\documentclass[sigconf]{acmart}

\usepackage{graphics,todonotes}
\usepackage{color}
\usepackage{booktabs} 
\usepackage{bm}
\usepackage{multirow}
\usepackage{diagbox,makecell}
\usepackage{threeparttable}

\AtBeginDocument{%
  \providecommand\BibTeX{{%
    \normalfont B\kern-0.5em{\scshape i\kern-0.25em b}\kern-0.8em\TeX}}}

\setcopyright{acmcopyright}
\copyrightyear{2022}
\acmYear{2022}
\acmDOI{10.1145/3503161.3548429}

\acmConference[ACMMM]{ACM International Conference on Multimedia}{October 10-14, 2022}{Lisbon, Portugal}
\newcommand{\etal}{\textit{et al.}}

\newcommand{\eg}{\textit{e.g.}, }

\newcommand{\ql}[1]{\textcolor{black}{#1}}
\newcommand{\zy}[1]{\textcolor{black}{#1}}

\begin{document}

\title{Immunofluorescence Capillary Imaging Segmentation: Cases Study}

\author{Runpeng Hou}
\email{12132626@mail.sustech.edu.cn}
\affiliation{%
  \institution{Southern University of Science and Technology}
  \city{Shenzhen}
  \state{Guangdong}
  \country{China}
}
\author{Ziyuan Ye}
\email{yezy2020@mail.sustech.edu.cn}
\affiliation{%
  \institution{Southern University of Science and Technology}
  \city{Shenzhen}
  \state{Guangdong}
  \country{China}
}
\author{Chengyu Yang}
\email{11930896@mail.sustech.edu.cn}
\affiliation{%
  \institution{Southern University of Science and Technology}
  \city{Shenzhen}
  \state{Guangdong}
  \country{China}
}
\author{Linhao Fu}
\email{11910321@mail.sustech.edu.cn}
\affiliation{%
  \institution{Southern University of Science and Technology}
  \city{Shenzhen}
  \state{Guangdong}
  \country{China}
}
\author{Chao Liu}
\email{liuc33@sustech.edu.cn}
\affiliation{%
  \institution{Southern University of Science and Technology}
  \city{Shenzhen}
  \state{Guangdong}
  \country{China}
}
\author{Quanying Liu}
\authornote{Quanying Liu is the corresponding authors}
\email{liuqy@sustech.edu.cn}
\affiliation{%
  \institution{SUSTech}
  \city{Shenzhen}
  \state{Guangdong}
  \country{China}
}


\begin{abstract}
  Nonunion is one of the challenges faced by orthopedics clinics for the technical difficulties and high costs in photographing interosseous capillaries. Segmenting vessels and filling capillaries are critical in understanding the obstacles encountered in capillary growth. However, existing datasets for blood vessel segmentation mainly focus on the large blood vessels of the body, and the lack of labeled capillary image datasets greatly limits the methodological development and applications of vessel segmentation and capillary filling.  Here, we present a benchmark dataset, named IFCIS-155, consisting of 155 2D capillary images with segmentation boundaries and vessel fillings annotated by biomedical experts, and 19 large-scale, high-resolution 3D capillary images. To obtain better images of interosseous capillaries, we leverage state-of-the-art immunofluorescence imaging techniques to highlight the rich vascular morphology of interosseous capillaries. We conduct comprehensive experiments to verify the effectiveness of the dataset and the benchmarking deep learning models (\eg UNet/UNet++ and the modified UNet/UNet++). Our work offers a benchmark dataset for training deep learning models for capillary image segmentation and provides a potential tool for future capillary research. The IFCIS-155 dataset and code are all publicly available at \url{https://github.com/ncclabsustech/IFCIS-55}.
\end{abstract}

\ccsdesc[500]{Computing methodologies~Image segmentation}



\keywords{IFCIS-155, immunofluorescence capillary, image segmentation, UNet.}

\maketitle

\section{Introduction}
The primary function of the vascular system is to deliver oxygen and glucose to metabolically active cells across the capillary bed, while concurrently removing CO$_2$ and other metabolic waste products. Arteries and veins can thus be viewed as structures that support capillary function~\cite{grutzendler2019cellular}. \ql{The health of capillaries largely determines the health of an organism.}
Vascular disorders and their downstream sequelae are responsible for more morbidity and mortality than any other category of human disease~\cite{bloodVessels}.  \ql{Vascular images characterize the morphology, distribution, and other important properties of blood vessels, which are closely related to vascular disorders. Therefore,} analyzing vascular images, especially capillary \ql{images}, plays a \ql{crucial} role in addressing the issue of clinical diagnosis and pathological analysis ~\cite{moccia2018blood}.

\zy{The interosseous capillary segmentation method can be divided into three main approaches:} manual segmentation, semi-automatic segmentation, and automatic segmentation.
Manual segmentation \zy{is a time-consuming and labor-intensive method}
which makes it impractical for \zy{real-world} applications \zy{with the growing demand of data}. 
Semi-automatic segmentation requires manually labeled vessel edge information, which makes the algorithm only provides limited convenience to the researcher due to the complex morphology of capillaries in interstitial healing tissue~\cite{gomariz2018quantitative}. \zy{The state-of-the-art automatic segmentation methods provide an effective and efficient way for immunofluorescence capillary imaging segmentation where 
most of them rely on threshold segmentation and surface rendering.}

\zy{The main challenges of immunofluorescence capillary imaging segmentation are the complexity and diversity of the objects' morphology, the large gap between the foreground and background, as well as the incomplete morphology of the vessel images. In recent years, developments in deep learning have opened up the possibility of end-to-end automatic segmentation of interosseous capillaries.
Consequently, we propose applying such methods to improve the performance of capillary segmentation in immunofluorescence imaging~\cite{gomariz2018quantitative}.}

\zy{However, to the best of our knowledge, none of the existing work has been applied to the segmentation of immunofluorescence imaging because of the lack of publicly available datasets.}
\zy{The current segmentation datasets have little contribution to our problem due to (1) imaging methods; (2) imaging environments; (3) vessel morphology. Therefore, transfer learning techniques and pretraining techniques are not suitable for the segmentation of immunofluorescence imaging.}

In this paper, we propose IFCIS-155 dataset: a high-quality, high-resolution immunofluorescence capillary imaging dataset \zy{for the segmentation task with well-annotations by experts. Furthermore, we adapt several state-of-the-art segmentation methods to tackle this problem.} The dataset contains 155 2D interosseous callus capillary images with 256$\times$256 pixels which contain rich capillary morphology, as well as expert annotated blood vessel repair images and blood vessel segmentation masks. \zy{Data are derived from 24 subjects who perform interosseous capillary sections.} We screened out 19 3D capillary fragments from 3D images of 3072$\times$1024$\times$20 pixels and divided them into 155 2D capillary images. 
\zy{The dataset included images depicting different morphological vessel segments, in addition to morphological vessel segments at different locations.}
\zy{The lack of training data prevents machine learning from being used to get better results from automatic capillary segmentation of interosseous capillaries.}
Here, we provide a publicly available interosseous capillary dataset and benchmark to fill the gap. 
\zy{Our experiments on IFCIS-155 data validated the feasibility of automatic deep learning segmentation for capillary segmentation.}

The main contributions of this study can be summarized as:
\begin{itemize}
    \item We propose a well-annotated interosseous callus capillary benchmark, IFCIS-155, which is specially designed for image segmentation in interosseous capillary scenes.
    
    \item We introduce a deep learning pipeline and streamline the research over the IFCIS-155 dataset with several state-of-the-art segmentation models, which could be a baseline reference for the IFCIS-155 dataset.
    
    \item We conduct extensive analyses on IFCIS-155 dataset using state-of-the-art models with different metrics, as well as performance comparisons with/without using image augmentation technique.

\end{itemize}

\begin{figure*}[htbp]
  \centering
  \includegraphics[width=\textwidth]{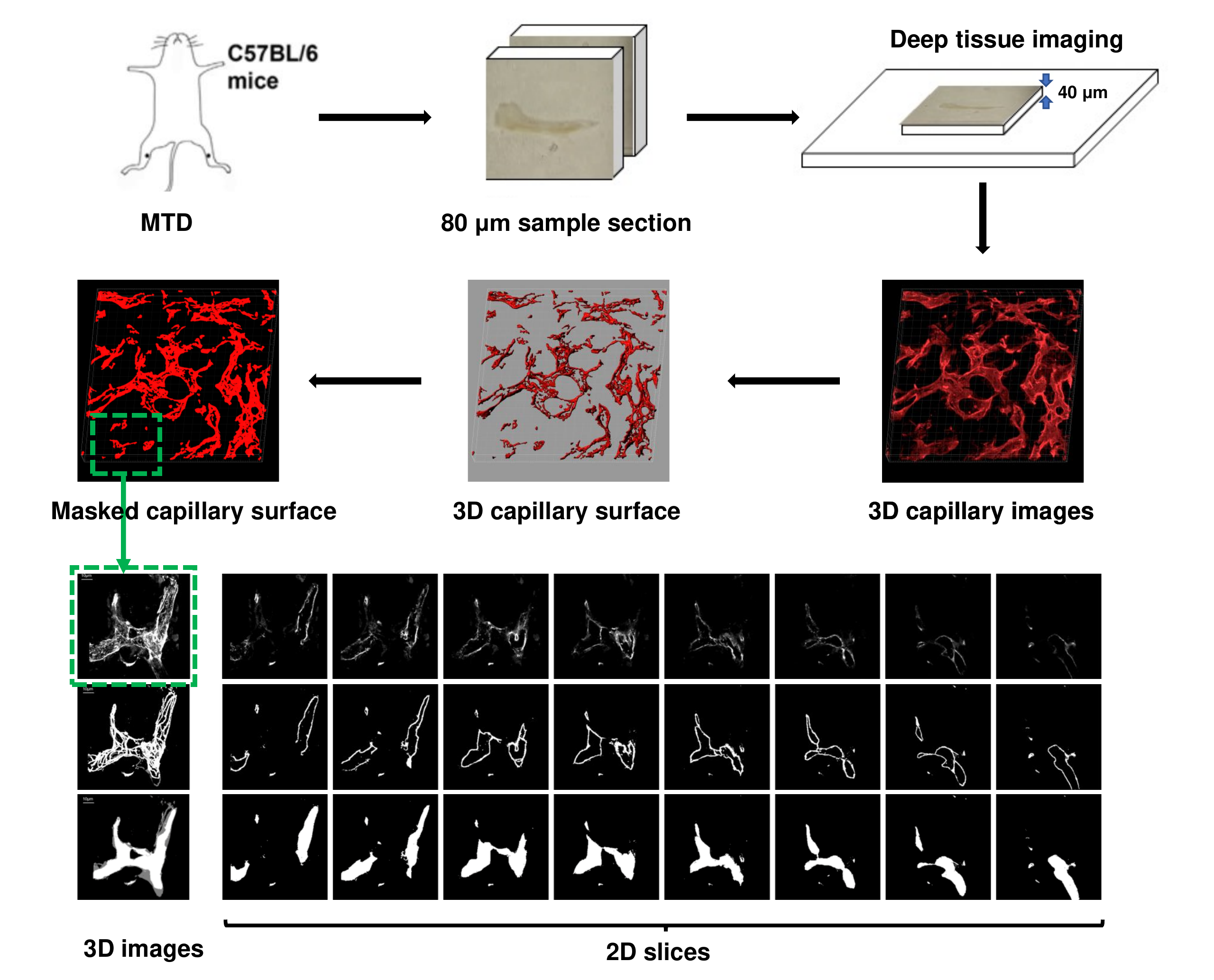}
  \caption{The acquisition process of the IFCIS-155 dataset. The first and second row of the figure shows the process of acquiring interstitial capillary images. We selected and annotated the 3D fragments of the interstitial capillary images of mice acquired during the experiment. The third row of the image shows the original and annotated sample image in IFCIS-155. The left of the third row is the vertical projection of the 3D image, and the right of the third row is the 2D slices selected for training and testing. From the first row to the third row of 2D slices are the original interosseous capillary images, the capillary boundary labels, and the capillary filling labels.}
  \label{fig:sample-data}
\end{figure*}

\section{Related Work}

\subsection{Datasets on Capillary}
Assessment of vascular characteristics plays an important role in various medical diagnoses\cite{saran2014role,sivaraj2016blood}. Several datasets of vessels with different morphologies have been proposed worldwide. There are relatively mature datasets for fundus capillary images, such as the STructured Analysis of the Retina (STARE) ~\cite{845178} dataset proposed by Hoover \etal, which includes hand-labeled ground truth segmentations of 20 images. Each image was digitized to produce a 605$\times$700 pixel image, 24 bits per pixel (standard RGB). In addition, the Digital Retinal Images for Vessel Extraction(DRIVE)~\cite{1282003} dataset is for retinal vessel segmentation proposed by J. Staal \textit{et al.} It consists of a total of JPEG 40 color fundus images; including 7 abnormal pathology cases. Each image resolution is 584*565 pixels with 
8 bits per color channel (3 channels). The images were obtained from a diabetic retinopathy screening program in the Netherlands. All images are fundus capillaries taken from the back and include manual segmentation and a circular field of view (FOV) mask.

Fundus imaging technology is mature and clear, however, there are many other forms of blood vessels in the human body. With the development of intrathoracic vascular imaging technology, the Intrapapilary capillary loops (IPCL)~\cite{garcia2020intrapapillary} dataset proposed by Garcia-Peraza-Herrera \textit{et al.} contains 68K binary labeled frames extracted from 114 patients videos for the detection of the mucosal layer of oesophagus lesions. In the imaging of interosseous capillaries, Ramasamy \textit{et al.} used a laser scanning confocal microscope to image interosseous capillaries in their experiments. Most of the current capillary datasets are different from in vivo capillary morphology (fundus) or labeling tasks (IPCL). To our knowledge, there is currently no publicly available annotated interosseous capillary image dataset.

\subsection{Blood Vessel Segmentation}
Vessel segmentation has received extensive attention in the past decade. Due to its importance and broad application prospects, many image segmentation algorithms have been applied in vessel segmentation. Although no general algorithm for vessel segmentation has yet been found, past research can provide many methods for reference. Many threshold-based image segmentation methods that show good performance in traditional image segmentation tasks, have the characteristics of fast calculation speed and high efficiency and have received extensive attention in the past. Such as the threshold segmentation method, histogram double peak method~\cite{iterative}, maximum entropy method, and Otsu method. Besides the above methods, the watershed algorithm, and region growing method are also applied to image segmentation tasks. While these traditional segmentations are highly efficient, with advances in vascular imaging, more and more vessels with complex shapes are being recorded. The accuracy of traditional segmentation algorithms is gradually unable to meet the requirements of blood vessel segmentation tasks in different scenarios.

With the rapid development of deep learning, it has gradually achieved state-of-the-art performance in more and more vision tasks, such as object detection, image classification, semantic segmentation, and object tracking~\cite{voulodimos2018deep}. Due to its data-driven features, the model can actively learn features from target data and complete specified tasks, which provides great convenience for complex vessel segmentation tasks. Deep learning models based on convolutional neural networks are widely used in image recognition and semantic segmentation. For example, the representative FCN~\cite{long2015fully},  DeepLab~\cite{chen2017deeplab}, UNet~\cite{U-Net} and other models all use convolutional layers for processing. These deep models extract different information at different network layers, and the prediction accuracy can be close to or even higher than that of human beings. Significant progress has been made for semantic segmentation of medical images in related work, for instance, FCN for organ segmentation ~\cite{ROTH201890}, SegNet for brain tumor segmentation~\cite{alqazzaz2019automated},  DeepLab for liver segmentation~\cite{tang2020two} and U-Net for cell segmentation, lesion segmentaition~\cite{U-Net,jaeger2020retina}. Recently, in the field of capillary imaging segmentation, deep learning methods also have been employed, such as fundus vessel segmentation~\cite{845178}, body surface capillary segmentation~\cite{8955887}, brain vessel segmentation~\cite{haft2019deep} and zebrafish embryo vessel segmentation~\cite{sun2022machine}. Among them, the UNet model, a symmetrical expansion path composed of an encoder and a decoder, has shown excellent results in medical image segmentation tasks. The encoder is used to obtain image features through convolution and pooling, and then the decoder is used for precise positioning. U-Net applies a skip-connection to the network to splice the feature map generated by the encoder to the up-sampled feature map of the decoder at each stage to learn the features lost in the encoder pooling process, which proved the possibility of end-to-end training from a small sample data with weighted loss. ~\cite{U-Net} After this, many UNet-based deep neural network has shown better results in some biomedical image segmentation tasks. ~\cite{U-Net++v2}

\begin{figure*}[h]
  \centering
  \includegraphics[width=\textwidth]{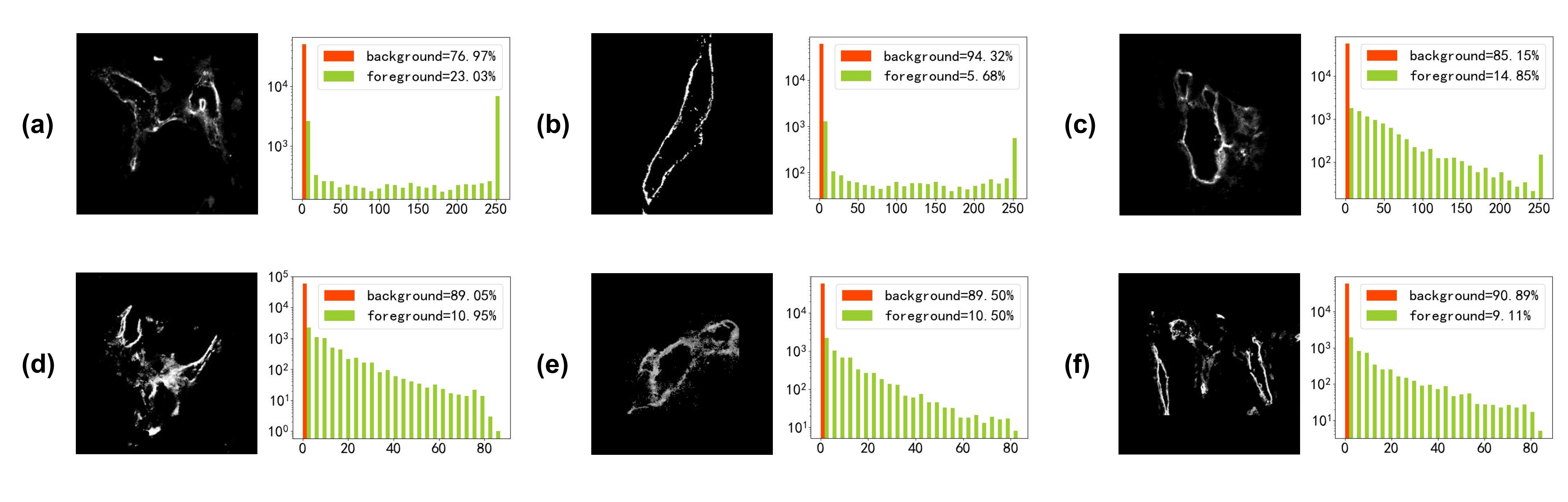}
  \caption{Histogram of pixel values. These pictures in black show six images selected from the training set, and histogram pictures show the pixel distribution in each image, where the green lines represent the foreground pixel and the red line represents the background pixel. The figure shows the proportion of foreground and background in the top right corner. The abscissa represents the  pixel value, and the ordinate corresponds to the number of times the pixel value appears in the image.}
  \label{fig:pixel-statistics}
\end{figure*}

\section{Dataset}

One of the main contributions of this paper is the Immunofluorescence Capillary Imaging Segmentation Dataset or IFCIS-155. Our principles in the design of IFCIS are:
\begin{itemize}
\item Get the clearest, most complete original image
\item Annotated images contain rich blood vessel morphology and accurate annotation information
\end{itemize}
Based on the above two principles, we collected and constructed our dataset in the following ways. The dataset was collected, annotated, and partitioned by a total of four experts. The collection process of the dataset meets ethical requirements and all in vivo animal protocols were approved by the Institutional Animal Care and Use Committee (certificate SUSTC-JY20190 427). 

Figure \ref{fig:sample-data} shows the image collection process and sample images.  The bilateral monocortical defects (MTD) were made in the tibia of C57BL/6 mice. This model consists of a 1-mm-diameter circular defect on the anterior medial surface of the tibia. Mice were euthanized on post-surgery day 10 and tibias were harvested and processed depending on the previous method ~\cite{liu2019mechanical}. 3D capillary images were obtained by using immunofluorescence staining of endomucin, which could label capillary in bone ~\cite{ramasamy2014endothelial}. And deep confocal laser technology has been applied to deep tissue imaging to scan the callus of the mouse 1mm single-cortical tibial defect model 10 days after surgery. Each initial 3D image is collected with Z-stacks of 40 $\mu m$ in thickness taken at a size of 3072×1024  pixels, x-y resolution of 0.624 $\mu m$ with z-step of 2 $\mu m$ in each initial image. The 3D surface of endomucin was created from deep tissue images by utilizing Imaris (version 7.1, Oxford Instruments, Switzerland), then masked. 3D images of interosseous capillary callus were acquired from 24 subjects, then we selected high-resolution 3D interosseous capillary images with the best signal brightness/sensitivity. Next, we cut the masked capillary blood vessel images of different morphology were cut into smaller 3D images. The image, which was manually filled according to the 3D capillary surface, was separated to create a 2D section. 

The dataset contains 155 sets of $256\times 256$ pixel 2D training images, vessel repair images, and vessel filling masks together with 19 3D unlabelled raw images cropped from 3D large-scale interosseous capillary images. As shown in Figure ~\ref{fig:pixel-statistics}, due e to the difference in the distribution of pixel values between images during deep immunofluorescence imaging, we performed image enhancement (contrast stretching) on the original image to facilitate model training.

\begin{figure*}[t]
\centering
\includegraphics[width=0.8\textwidth]{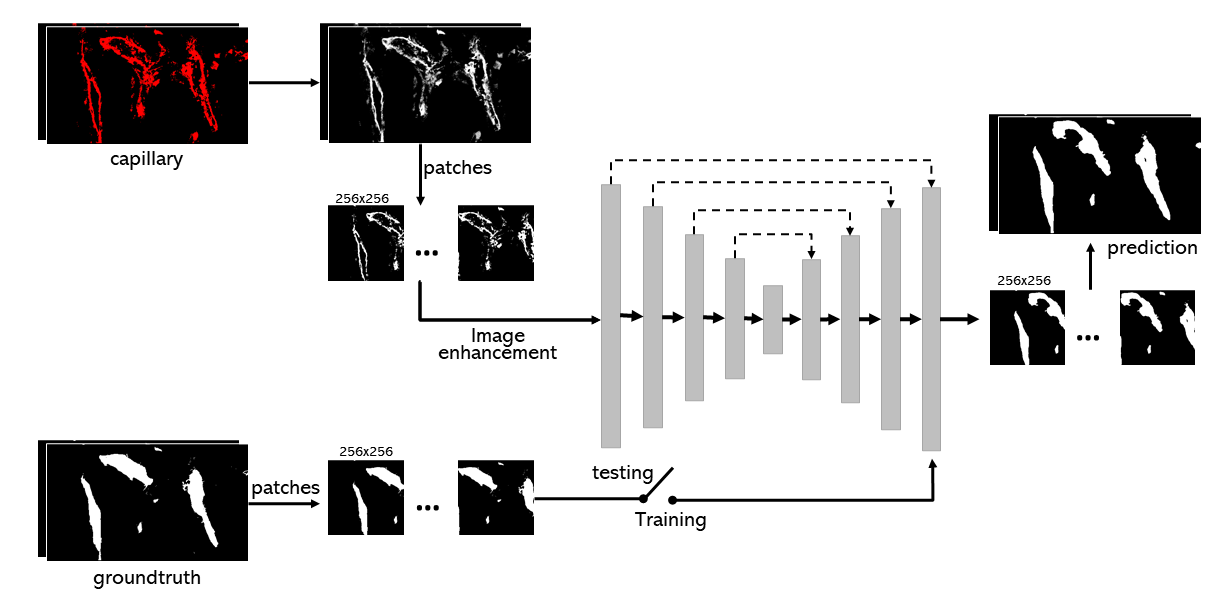}
\caption{Slicing and Deep learning pipeline. When the original image does not meet the size of the input image, the sliding window method will be used to cut the original image into several 256x256 images and then input them into the model for prediction. After the output result is obtained, the predicted image is reorganized back to the original image size as the predicted result.}
\label{fig_network_architecture}
\end{figure*}

\section{Methods for capillary segmentation on IFCIS-155}
In this section, we discuss a number of models used in the IFCIS-155 dataset. We combine multiple data augmentation, preprocessing, and baseline methods to analyze the performance of existing models on the interosseous capillary segmentation task, as well as the main challenges. Then we propose the pipeline for large-scale image segmentation using deep learning methods together with the objective functions and the evaluation metrics used during training.

\subsection{Data Augmentation}

Data augmentation is an operation that generates a new image while preserving the intrinsic characteristics (\eg label) of the original image. In addition to increasing the number of training sets, appropriate data augmentation strategies can improve the robustness of the training datasets and the generalization ability of the model. In particular, when the data size is limited, data augmentation is essential for model training. Commonly used data augmentation methods for natural images include rotation, translation, scaling, cutting, masking, symmetry, etc. In order to preserve the scale information of immunofluorescence imaging, we mainly used rotation, translation, shearing, and horizontal flip in our experiments. This strategy, which brings an increment of about 10\% on prediction accuracy, allows us to train a model with better robustness under the condition of limited data.

\subsection{Slicing and Preprocessing}

As mentioned above, we perform image augmentation on the input data to make the dataset more conducive to model training. All image pixels are scaled uniformly until the maximum pixel value reaches the upper limit. This does not affect conventionally captured pictures but improves the observability of darker images. In addition, we randomly divided the data set into five folds of the same number of images and performed five-fold cross-validation. Except for the fold for testing, the other four folds can be used for data augmentation to increase the robustness of the model.

Since the size of the original interosseous capillary images is usually much larger than the test data, directly inputting large-scale images brings a large memory and computational burden to the network. When segmenting the original large-scale capillary images, we use sliding windows to segment the large-scale images into standard block sizes for prediction. Blocks in adjacent positions overlap. When the prediction results are aggregated, the overlapping parts of the prediction results are determined by a joint vote of multiple blocks to reduce the error in the boundary area.

\begin{figure*}[htbp]
  \centering
  \includegraphics[width=\textwidth]{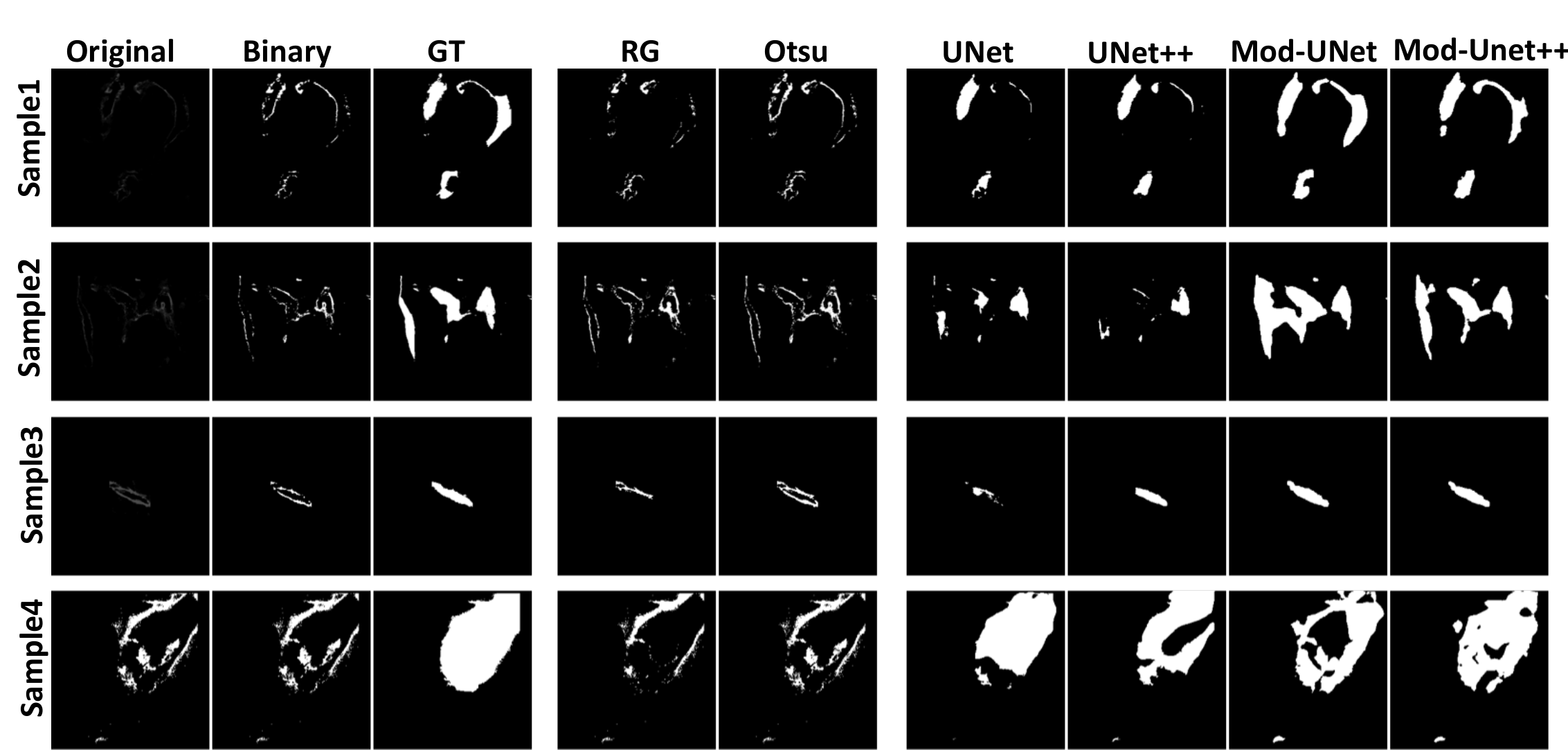}
  \caption{Sample results of immunofluorescence capillary imaging segmentation task. The left three columns are the original image, enhanced image, and the ground truth. The two columns in the middle are the segmentation results of the region growth method and the Otsu method. The last four columns on the right are the prediction results of the deep model.}
  \label{fig:prediction-result}
\end{figure*}

\subsection{Deep-learning Pipelines}
\subsubsection{Network Architecture}\

In this paper, we mainly use UNet and UNet++ as the baseline methods for testing and analysis of the IFCIS-155 dataset. Figure 3 illustrates the slicing method and deep learning pipeline. During the training process, we input 256×256 training data and ground truth. If the shape of the original image is not 256×256, the sliding window will be used to segment the original image and obtain the transformed image with 256×256. Then, the image enhancement (contrast stretching) technique is implemented to adjust the distribution of pixel values and input them into the prediction model. There are no data leakage issues when splitting datasets.

\textbf{UNet} has received extensive attention since its appearance in 2015 and has been frequently applied to medical image segmentation tasks. UNet has an encoder-decoder part, which enables UNet to extract image features and restore segmented masks. Different from the traditional encoder-decoder network, UNet adds skip-connection, which restores some spatial features lost during feature extraction and greatly improves the prediction results. However, the skip-connection structure of UNet is relatively simple, and may not work well when performing complex object segmentation tasks. 

\textbf{UNet++} proposed in~\cite{zhou2018unet++} improved the skip connection form based on UNet. UNet++ with a richer skip-connection structure is a multi-scale, high-density convolutional network. And UNet++ has shown better performance on cell segmentation and brain tumor segmentation task~\cite{U-Net++v2}.

\subsubsection{Loss Function and Evaluation Criterion }\

\textbf{BCE, binary cross-entropy}
Since the model only needs to segment two types of pixels for a single recognition, we use binary cross-entropy as one of the baseline loss functions:

$$L=-\frac{1}{N}\sum_{x\in X}[y\cdot ln\hat y+(1-y)\cdot ln(1-\hat y)]$$
where $N$ represents the number of samples, $X$ represents the set of all pixels, $y$ represents the label, and $\hat y$ represents the predicted value. We compute the binary cross-entropy of all training samples in each batch pixel by pixel.

\textbf{Joint loss}
Joint loss is obtained by adding a cross entropy and a Dice loss:
 $$L_{Joint}=L_{Dice}+L_{CE}$$
 $$L_{Dice}=-\frac{2}{|K|}\sum_{k\in K}\frac{\sum_{i\in I} u_i^kv_i^k}{\sum_{i \in I}u_i^k+\sum_{i\in I}v_i^k}$$
 where $u$ is the softmax output of the network and $v$ is a one hot encoding
  of the ground truth segmentation map. Both $u$ and $v$ have shape $I \times K$ with
  $i \in I$ being the number of pixels in the training patch/batch and $k \in K$ being
  the classes.


To evaluate the performance of baseline methods, several evaluation metrics were applied, including accuracy (ACC), recall, precision, Dice coefficient, and Intersection over Union (IoU). ACC is used to measure the proportion of correct predictions. Precision refers to the proportion of true positive samples to all positive samples. Recall measures the proportion of positive samples that are correctly identified. And Dice is used to balance precision and recall. IoU gauges the similarity of ground truth and predicted mask. These metrics are defined as:
$$ACC=\frac{TP+TN}{TP+TN+FP+FN}$$
$$Recall=\frac{TP}{TP+FN}$$
$$Precision=\frac{TP}{TP+FP}$$
$$Dice=\frac{2\times TP}{2TP+FP+FN}$$
$$IoU=\frac{TP}{TP+FP+TN}$$
where true positive ($TP$) and true negative ($TN$) refers to the number of pixels which was correctly predicted for the positive class and negative class. False positive($FP$) and false negative($FN$) represent the number of incorrect pixels which were predicted for the positive class and negative class. 



\section{Experimental Evaluation}

\subsection{Experimental Settings}
In this section, we evaluate and present the segmentation results of two traditional segmentation algorithms as well as four deep learning segmentation models. RTX 3080 and Tesla V100 GPU were used during training. The segmentation results are expressed as the average of five training results. Five-fold validation tests were used in each process of training a deep learning model. The parameters of the deep learning models were initialized with he normal~\cite{he2015delving}, and each model has an initial learning rate of $10^{-4}$. We trained each model for 100 epochs with gradient descent using the Adam optimizer. We tested UNet and UNet++ with different loss functions and add the models with the best prediction result to the experiment for comparison, named modified-UNet and modified-UNet++.


\subsection{Performance Analysis}
\begin{table*}[htbp]
	\centering
	\caption{Quantitative results across 5-fold cross-validation on capillary images. }
	\begin{tabular}{c|ccc|ccc|ccc|ccc|ccc}
		\toprule
		\multirow{1}*{\diagbox{Models}{Metrics}} & \multicolumn{3}{c|}{Dice}  & \multicolumn{3}{c|}{IoU} & \multicolumn{3}{c|}{Precision} & \multicolumn{3}{c|}{Recall} & \multicolumn{3}{c}{Acc}\\ 
		& AVG & STD & BEST & AVG & STD & BEST & AVG & STD & BEST & AVG & STD & BEST & AVG & STD & BEST\\
		\midrule
		Otsu & 16.44 & - & - & 10.69 & - & - & 39.83 & - & - & 11.07 & - & - & - & - & - \\
		RG & 12.79 & - & - & 7.85 & - & - & 35.04 & - & - & 8.60 & - & - & - & - & -\\
		Unet & 45.54 & 34.59 & 88.30 & 35.51 & 32.36 & 77.13 & 40.59 & 41.31 & 84.85 & 55.81 & 45.58 & 94.81 & 95.63 & 2.83 & 97.99\\
		Unet$++$ & 64.91 & 11.99 & 88.83 & 47.06 & 13.61 & 74.22 & 57.62 & 4.84 & 91.07 & 91.56 & 2.40 & 94.37 & 97.76 & 0.50 & 98.30\\
		\bottomrule
	\end{tabular}
	\label{table:predictionDice}
	\begin{tablenotes}[flushleft]
     \item Note: AVG, STD and BEST denote average result, standard deviation and best result, respectively.
   \end{tablenotes}
\end{table*}

\begin{table*}[htbp]
	\centering
	\caption{Quantitative results across 5-fold cross-validation on capillary images with different loss. }~\label{tab:tab4name}
	\begin{tabular}{c|c|ccc|ccc|ccc|ccc|ccc}
		\toprule
		\multirow{2}*{Models} &
		\multirow{2}*{Loss} &
		\multicolumn{3}{c|}{Dice}  & \multicolumn{3}{c|}{IoU} & \multicolumn{3}{c|}{Precision} & \multicolumn{3}{c|}{Recall} & \multicolumn{3}{c}{Acc}\\ 
		& & AVG & STD & BEST & AVG & STD & BEST & AVG & STD & BEST & AVG & STD & BEST & AVG & STD & BEST\\
		\midrule
		Unet & BCE & 45.54 & 34.59 & 88.30 & 35.51 & 32.36 & 77.13 & 40.59 & 41.31 & 84.85 & 55.81 & 45.58 & 94.81 & 95.63 & 2.83 & 97.99\\
		Unet & Joint & 70.43 & 24.93 & 89.78 & 57.33 & 26.10 & 79.03 & 66.45 & 3.34 & 91.81 & 82.87 & 2.17 & 95.00 & 98.18 & 0.19 & 98.45\\
		Unet++ & BCE & 64.91 & 11.99 & 88.83 & 47.06 & 13.61 & 74.22 & 57.62 & 4.84 & 91.07 & 91.56 & 2.40 & 94.37 & 97.76 & 0.50 & 98.30\\
		Unet++ & Joint & 76.12 & 25.22 & 89.81 & 64.41 & 25.45 & 79.52 & 72.41 & 28.81 & 88.41 & 91.89 & 3.24 & 95.63 & 98.34 & 0.19 & 98.63\\
		\bottomrule
	\end{tabular}
	\begin{tablenotes}[flushleft]
     \item Note: AVG, STD and BEST denote average result, standard deviation and best result, respectively.
   \end{tablenotes}
\end{table*}

The complete restoration of the vascular shape can be used to study the mechanism of biological phenomena, so we mainly focus on segmenting the blood vessel region, which is named the immunofluorescence capillary imaging segmentation task. We tested different models on IFCIS-155 dataset, hoping to restore as many original capillary shapes as possible. Figure \ref{fig:prediction-result} shows the prediction results of six different segmentation methods for four sample images. The first three columns (left) in Figure \ref{fig:prediction-result} show the original images of the four sample images (Original), the images obtained after binarization (Binary), and the ground truth (GT) of the blood vessel annotations. 

\textbf{Experiment 1}: We first tested IFCIS-155 using different image segmentation methods, and evaluated the performance of traditional methods and deep learning methods on the capillary segmentation task. We also quantitatively analyze and visualize the results. The fourth and fifth columns of images in Figure \ref{fig:prediction-result} are the predicted images obtained using the region growing method and the Otsu method, respectively. Since most of the center of capillary in original images are weakly or not imaging, there is no clear benefit for solving immunofluorescence capillary image segmentation tasks with the traditional methods. Therefore, we use the deep learning models mentioned above for segmentation, and depict the prediction results in the sixth to ninth columns of Figure \ref{fig:prediction-result} -- prediction results of UNet, UNet++, Modified-UNet (Mod-UNet), Modified-UNet++ (Mod-UNet++). We use the Dice coefficient to quantify the prediction accuracy of these deep learning models in Table \ref{table:predictionDice}. Table \ref{table:predictionDice} compares the prediction results of four algorithms among five measures. On average, there is a significant difference between the prediction results of deep learning models and traditional segmentation algorithms. Not surprisingly, the deep learning model has more than 30\% performance improvement on Dice and IoU compared to traditional segmentation algorithms. In experiment 1, UNet++ showed the best performance in all indicators.

\textbf{Experiment 2}: Due to the large difference of pixels between the foreground and background of capillary images shown in Figure \ref{fig:pixel-statistics}, we tested the effects of different loss functions in the deep learning model and carried out a quantitative analysis. We use joint loss in experiment 2 to compare with the original model. As can be seen from Table\ref{tab:tab4name}, the joint loss brings a certain improvement to the prediction accuracy of UNet and UNet++. The improvement reaches 25\% in Dice of UNet and 12\% in Dice of UNet++. The themes identified in these comparison shows that joint loss is more suitable for capillary segmentation tasks

\textbf{Experiment 3}: To explore the effect of data augmentation on training performance, we performed the data augmentation described in section 4.1 on FICIS-155, applied it to train the deep learning model, and quantitatively analyzed the results. The purpose of experiment 3 was to explore the effect of data augmentation on training performance. Experiment 3 contains two sets of comparative models. We input the augmented and original data into the same model to compare the prediction performance after model training. Five metrics were used to evaluate the prediction performance in simple statistical analysis. Table \ref{table:experiment2} shows the impact of data augmentation on performance. It is apparent from this table that, in both UNet and UNet++, data augmentation brings a significant improvement to evaluation metrics. The improvement reaches 25\% in Dice of  UNet and 7\% in UNet++. Table \ref{tab:tab3name} also presents an increase in Dice even with joint loss. In evaluating IoU and Precision, the model after data augmentation also has different degrees of improvement. Interestingly,  the standard deviation of two pairs of comparison models was observed to have a large gap. Such results indicate that data augmentation can be applied to IFCIS-155, although the model has a certain predictive ability for small sample data.

\begin{table*}[htbp]
	\centering
	\caption{Quantitative results across 5-fold cross-validation on capillary images with/without data augmentation. }
	
	\begin{tabular}{c|c|ccc|ccc|ccc|ccc|ccc}
		\toprule
		\multirow{2}*{Models} &
		\multirow{2}*{\makecell[c]{Data\\Aug?}} &
		\multicolumn{3}{c|}{Dice}  & \multicolumn{3}{c|}{IoU} & \multicolumn{3}{c|}{Precision} & \multicolumn{3}{c|}{Recall} & \multicolumn{3}{c}{Acc}\\ 
		& & AVG & STD & BEST & AVG & STD & BEST & AVG & STD & BEST & AVG & STD & BEST & AVG & STD & BEST\\
		\midrule
		Unet &  $\times$ &  45.54 & 34.59  & 88.30  & 35.51  & 32.36  & 77.13  & 40.59  & 41.31  & 84.85  & 55.81  & 45.58  & 94.81  & 95.63  & 2.83  & 97.99 \\
		Unet & $\checkmark$ & 70.15 & 3.52  & 75.70  & 49.97  & 4.20  & 54.38  & 60.10  & 5.87  & 70.85  & 85.18  & 2.97  & 88.59  & 96.21  & 0.82  & 97.50  \\
		Unet++ &  $\times$ & 64.91  & 11.99  & 88.83  & 47.06  & 13.61  & 74.22  & 57.62  & 4.84  & 91.07  & 91.56  & 2.40  & 94.37  & 97.76  & 0.50  & 98.30 \\
		Unet++ & $\checkmark$  & 71.77  & 10.47  & 80.29  & 52.61  & 10.07  & 60.40  & 67.80  & 6.15  & 73.28  & 77.08  & 15.23  & 88.82  & 96.78  & 0.78  & 97.30 \\
		\bottomrule
	\end{tabular}
	\begin{tablenotes}[flushleft]
     \item Note: AVG, STD and BEST denote average result, standard deviation and best result, respectively. ``Data Aug'' indicates the experiments are conducted with data augmentation.
   \end{tablenotes}
	\label{table:experiment2}
\end{table*}


\begin{table*}[htbp]
	\centering
	\caption{Quantitative results across 5-fold cross-validation on capillary images segmentation task using joint loss \zy{with/without} data augmentation.}~\label{tab:tab3name}
	\setlength{\tabcolsep}{4.3pt}
	\begin{tabular}{c|c|c|ccc|ccc|ccc|ccc|ccc}
		\toprule
		\multirow{2}*{Models} &
		\multirow{2}*{\makecell[c]{Data\\Aug?}} &
		\multirow{2}*{Loss} &
		\multicolumn{3}{c|}{Dice}  & \multicolumn{3}{c|}{IoU} & \multicolumn{3}{c|}{Precision} & \multicolumn{3}{c|}{Recall} & \multicolumn{3}{c}{Acc}\\ 
		& & &AVG & STD & BEST & AVG & STD & BEST & AVG & STD & BEST & AVG & STD & BEST & AVG & STD & BEST\\
		\midrule
		Unet & $\times$ & Joint&70.43 & 24.93 & 89.78 & 57.33 & 26.10 & 79.03 & 66.45 & 3.34 & 91.81 & 82.87 & 2.17 & 95.00 & 98.18 & 0.19 & 98.45\\
		Unet & $\checkmark$ & Joint&84.18 & 1.90 & 87.87 & 69.77 & 1.99 & 72.10 & 78.99 & 3.59 & 85.29 & 90.39 & 3.07 & 93.01 & 97.73 & 0.36 & 98.30\\ 
		Unet++ & $\times$ & Joint&76.12 & 25.22 & 89.81 & 64.41 & 25.45 & 79.52 & 72.41 & 28.81 & 88.41 & 91.89 & 3.24 & 95.63 & 98.34 & 0.19 & 98.63\\
		Unet++ & $\checkmark$ & Joint&85.71 & 2.52 & 88.30 & 72.57 & 2.26 & 75.60 & 79.01 & 5.76 & 86.36 & 91.78 & 1.33 & 93.39 & 97.63 & 0.61 & 98.14\\
		\bottomrule
	\end{tabular}
	\begin{tablenotes}[flushleft]
     \item Note: AVG, STD and BEST denote average result, standard deviation and best result, respectively. ``Data Aug'' indicates the experiments are conducted with data augmentation.
   \end{tablenotes}
\end{table*}


\section{Discussion}
Previous studies have shown that the vascular morphology and structure information of interosseous capillaries can help study bone development, remodeling, and homeostasis. However, the segmentation of in vivo blood vessels mainly relies on manual annotation or traditional methods to repair the vessel morphology. Manual annotation is time-consuming and tedious, and the traditional methods are not scalable to the increased data size. When provided with a large amount of effective data and labels, deep learning models have achieved promising performance on surface vessel image classification tasks  and fundus image segmentation tasks. At present, there are relatively sufficient studies on fundus capillaries, which can be combined with machine learning models for fundus blood vessel segmentation and diabetic retinopathy screening. However, due to the lack of publicly available in vivo capillary images for training, the task of in vivo capillary image segmentation remains cumbersome and complex. In this study, we presented the IFCIS-155 dataset, which includes hand-annotated ground-truth segmentation labels and inpainted capillary labels for 155 images. our experiments show that using augmented data from IFCIS-155 deep learning models can achieve better segmentation performance. This dataset contains images of interosseous capillaries of different shapes, which can be extended to train models for other in vivo blood vessel segmentation tasks and have broad applications in the future.

The acquisition and labeling of medical image data are difficult and time-consuming. Different from the natural image segmentation tasks, the medical image segmentation tasks often have limited data.  UNet and related models have been widely used in previous medical image segmentation tasks since it simply builds an encoder and a decoder with skip connection, which provides efficient information flow and shows excellent performance in describing and segmenting objects of complex shapes. UNet++ retains the advantages of skip connection from UNet and introduces a more comprehensive connection. These models have been used in the past for small-sample medical image segmentation tasks. Moreover, it has also shown good adaptability in IFCIS-155. Similar to the previous medical small sample segmentation task, data augmentation also improves the robustness of the model in IFCIS-155, which brings improvements to the prediction effect.

\section{Conclusion}
In this paper, we mainly introduce a challenging, high-resolution benchmark dataset for interosseous capillary segmentation, IFCIS-155, which fills the gap in publicly annotated interosseous capillary datasets. To dissect the IFCIS-155 and facilitate future research on capillary segmentation, we also conduct an extensive evaluation of two baselines of the IFCIS-155 challenge, which can be a reference for future work. The unsatisfying performance on the IFCIS-155 dataset suggests that more attention should be paid to various types of real-world capillary segmentation. In the future, we plan to extend the IFCIS-155 dataset to relevant tasks, such as semi-supervised/unsupervised or few-shot capillary segmentation tasks. We hope that the IFCIS-155 dataset can help to push the research to the capillary segmentation task.

\section{Acknowledgment}
This work was funded in part by the National Key Research and Development Program of China (2021YFF1200800), National Natural Science Foundation of China (62001205),  Shenzhen-Hong Kong-Macao Science and Technology Innovation Project (SGDX2020110309280100), Guangdong Natural Science Foundation Joint Fund (2019A1515111038), Shenzhen Science and Technology Innovation Committee (20200925155957004, KCXFZ2020122117340001), Shenzhen Key Laboratory of Smart Healthcare Engineering (ZDSYS20200811144003009).

\bibliographystyle{ACM-Reference-Format}
\bibliography{references.bib}


\begin{thebibliography}{26}


\ifx \showCODEN    \undefined \def \showCODEN     #1{\unskip}     \fi
\ifx \showDOI      \undefined \def \showDOI       #1{#1}\fi
\ifx \showISBNx    \undefined \def \showISBNx     #1{\unskip}     \fi
\ifx \showISBNxiii \undefined \def \showISBNxiii  #1{\unskip}     \fi
\ifx \showISSN     \undefined \def \showISSN      #1{\unskip}     \fi
\ifx \showLCCN     \undefined \def \showLCCN      #1{\unskip}     \fi
\ifx \shownote     \undefined \def \shownote      #1{#1}          \fi
\ifx \showarticletitle \undefined \def \showarticletitle #1{#1}   \fi
\ifx \showURL      \undefined \def \showURL       {\relax}        \fi
\providecommand\bibfield[2]{#2}
\providecommand\bibinfo[2]{#2}
\providecommand\natexlab[1]{#1}
\providecommand\showeprint[2][]{arXiv:#2}

\bibitem[\protect\citeauthoryear{??}{ite}{1978}]%
        {iterative}
 \bibinfo{year}{1978}\natexlab{}.
\newblock \showarticletitle{Picture Thresholding Using an Iterative Selection
  Method}.
\newblock \bibinfo{journal}{\emph{IEEE Transactions on Systems, Man, and
  Cybernetics}} \bibinfo{volume}{8}, \bibinfo{number}{8}
  (\bibinfo{year}{1978}), \bibinfo{pages}{630--632}.
\newblock
\urldef\tempurl%
\url{https://doi.org/10.1109/TSMC.1978.4310039}
\showDOI{\tempurl}


\bibitem[\protect\citeauthoryear{Alqazzaz, Sun, Yang, and Nokes}{Alqazzaz
  et~al\mbox{.}}{2019}]%
        {alqazzaz2019automated}
\bibfield{author}{\bibinfo{person}{Salma Alqazzaz}, \bibinfo{person}{Xianfang
  Sun}, \bibinfo{person}{Xin Yang}, {and} \bibinfo{person}{Len Nokes}.}
  \bibinfo{year}{2019}\natexlab{}.
\newblock \showarticletitle{Automated brain tumor segmentation on multi-modal
  MR image using SegNet}.
\newblock \bibinfo{journal}{\emph{Computational Visual Media}}
  \bibinfo{volume}{5}, \bibinfo{number}{2} (\bibinfo{year}{2019}),
  \bibinfo{pages}{209--219}.
\newblock


\bibitem[\protect\citeauthoryear{Chen, Papandreou, Kokkinos, Murphy, and
  Yuille}{Chen et~al\mbox{.}}{2017}]%
        {chen2017deeplab}
\bibfield{author}{\bibinfo{person}{Liang-Chieh Chen}, \bibinfo{person}{George
  Papandreou}, \bibinfo{person}{Iasonas Kokkinos}, \bibinfo{person}{Kevin
  Murphy}, {and} \bibinfo{person}{Alan~L Yuille}.}
  \bibinfo{year}{2017}\natexlab{}.
\newblock \showarticletitle{Deeplab: Semantic image segmentation with deep
  convolutional nets, atrous convolution, and fully connected crfs}.
\newblock \bibinfo{journal}{\emph{IEEE transactions on pattern analysis and
  machine intelligence}} \bibinfo{volume}{40}, \bibinfo{number}{4}
  (\bibinfo{year}{2017}), \bibinfo{pages}{834--848}.
\newblock


\bibitem[\protect\citeauthoryear{Garc{\'\i}a-Peraza-Herrera, Everson, Lovat,
  Wang, Wang, Haidry, Stoyanov, Ourselin, and
  Vercauteren}{Garc{\'\i}a-Peraza-Herrera et~al\mbox{.}}{2020}]%
        {garcia2020intrapapillary}
\bibfield{author}{\bibinfo{person}{Luis~C Garc{\'\i}a-Peraza-Herrera},
  \bibinfo{person}{Martin Everson}, \bibinfo{person}{Laurence Lovat},
  \bibinfo{person}{Hsiu-Po Wang}, \bibinfo{person}{Wen~Lun Wang},
  \bibinfo{person}{Rehan Haidry}, \bibinfo{person}{Danail Stoyanov},
  \bibinfo{person}{S{\'e}bastien Ourselin}, {and} \bibinfo{person}{Tom
  Vercauteren}.} \bibinfo{year}{2020}\natexlab{}.
\newblock \showarticletitle{Intrapapillary capillary loop classification in
  magnification endoscopy: open dataset and baseline methodology}.
\newblock \bibinfo{journal}{\emph{International journal of computer assisted
  radiology and surgery}} \bibinfo{volume}{15}, \bibinfo{number}{4}
  (\bibinfo{year}{2020}), \bibinfo{pages}{651--659}.
\newblock


\bibitem[\protect\citeauthoryear{Gomariz, Helbling, Isringhausen, Suessbier,
  Becker, Boss, Nagasawa, Paul, Goksel, Sz{\'e}kely, et~al\mbox{.}}{Gomariz
  et~al\mbox{.}}{2018}]%
        {gomariz2018quantitative}
\bibfield{author}{\bibinfo{person}{Alvaro Gomariz}, \bibinfo{person}{Patrick~M
  Helbling}, \bibinfo{person}{Stephan Isringhausen}, \bibinfo{person}{Ute
  Suessbier}, \bibinfo{person}{Anton Becker}, \bibinfo{person}{Andreas Boss},
  \bibinfo{person}{Takashi Nagasawa}, \bibinfo{person}{Gr{\'e}gory Paul},
  \bibinfo{person}{Orcun Goksel}, \bibinfo{person}{G{\'a}bor Sz{\'e}kely},
  {et~al\mbox{.}}} \bibinfo{year}{2018}\natexlab{}.
\newblock \showarticletitle{Quantitative spatial analysis of
  haematopoiesis-regulating stromal cells in the bone marrow microenvironment
  by 3D microscopy}.
\newblock \bibinfo{journal}{\emph{Nature communications}} \bibinfo{volume}{9},
  \bibinfo{number}{1} (\bibinfo{year}{2018}), \bibinfo{pages}{1--15}.
\newblock


\bibitem[\protect\citeauthoryear{Grutzendler and Nedergaard}{Grutzendler and
  Nedergaard}{2019}]%
        {grutzendler2019cellular}
\bibfield{author}{\bibinfo{person}{Jaime Grutzendler} {and}
  \bibinfo{person}{Maiken Nedergaard}.} \bibinfo{year}{2019}\natexlab{}.
\newblock \showarticletitle{Cellular control of brain capillary blood flow: in
  vivo imaging veritas}.
\newblock \bibinfo{journal}{\emph{Trends in neurosciences}}
  \bibinfo{volume}{42}, \bibinfo{number}{8} (\bibinfo{year}{2019}),
  \bibinfo{pages}{528--536}.
\newblock


\bibitem[\protect\citeauthoryear{Haft-Javaherian, Fang, Muse, Schaffer,
  Nishimura, and Sabuncu}{Haft-Javaherian et~al\mbox{.}}{2019}]%
        {haft2019deep}
\bibfield{author}{\bibinfo{person}{Mohammad Haft-Javaherian},
  \bibinfo{person}{Linjing Fang}, \bibinfo{person}{Victorine Muse},
  \bibinfo{person}{Chris~B Schaffer}, \bibinfo{person}{Nozomi Nishimura}, {and}
  \bibinfo{person}{Mert~R Sabuncu}.} \bibinfo{year}{2019}\natexlab{}.
\newblock \showarticletitle{Deep convolutional neural networks for segmenting
  3D in vivo multiphoton images of vasculature in Alzheimer disease mouse
  models}.
\newblock \bibinfo{journal}{\emph{PloS one}} \bibinfo{volume}{14},
  \bibinfo{number}{3} (\bibinfo{year}{2019}), \bibinfo{pages}{e0213539}.
\newblock


\bibitem[\protect\citeauthoryear{Hariyani, Eom, and Park}{Hariyani
  et~al\mbox{.}}{2020}]%
        {8955887}
\bibfield{author}{\bibinfo{person}{Yuli~Sun Hariyani}, \bibinfo{person}{Heesang
  Eom}, {and} \bibinfo{person}{Cheolsoo Park}.}
  \bibinfo{year}{2020}\natexlab{}.
\newblock \showarticletitle{DA-Capnet: Dual Attention Deep Learning Based on
  U-Net for Nailfold Capillary Segmentation}.
\newblock \bibinfo{journal}{\emph{IEEE Access}}  \bibinfo{volume}{8}
  (\bibinfo{year}{2020}), \bibinfo{pages}{10543--10553}.
\newblock
\urldef\tempurl%
\url{https://doi.org/10.1109/ACCESS.2020.2965651}
\showDOI{\tempurl}


\bibitem[\protect\citeauthoryear{He, Zhang, Ren, and Sun}{He
  et~al\mbox{.}}{2015}]%
        {he2015delving}
\bibfield{author}{\bibinfo{person}{Kaiming He}, \bibinfo{person}{Xiangyu
  Zhang}, \bibinfo{person}{Shaoqing Ren}, {and} \bibinfo{person}{Jian Sun}.}
  \bibinfo{year}{2015}\natexlab{}.
\newblock \showarticletitle{Delving deep into rectifiers: Surpassing
  human-level performance on imagenet classification}. In
  \bibinfo{booktitle}{\emph{Proceedings of the IEEE international conference on
  computer vision}}. \bibinfo{pages}{1026--1034}.
\newblock


\bibitem[\protect\citeauthoryear{Hoover, Kouznetsova, and Goldbaum}{Hoover
  et~al\mbox{.}}{2000}]%
        {845178}
\bibfield{author}{\bibinfo{person}{AD Hoover}, \bibinfo{person}{Valentina
  Kouznetsova}, {and} \bibinfo{person}{Michael Goldbaum}.}
  \bibinfo{year}{2000}\natexlab{}.
\newblock \showarticletitle{Locating blood vessels in retinal images by
  piecewise threshold probing of a matched filter response}.
\newblock \bibinfo{journal}{\emph{IEEE Transactions on Medical imaging}}
  \bibinfo{volume}{19}, \bibinfo{number}{3} (\bibinfo{year}{2000}),
  \bibinfo{pages}{203--210}.
\newblock


\bibitem[\protect\citeauthoryear{Jaeger, Kohl, Bickelhaupt, Isensee, Kuder,
  Schlemmer, and Maier-Hein}{Jaeger et~al\mbox{.}}{2020}]%
        {jaeger2020retina}
\bibfield{author}{\bibinfo{person}{Paul~F Jaeger}, \bibinfo{person}{Simon~AA
  Kohl}, \bibinfo{person}{Sebastian Bickelhaupt}, \bibinfo{person}{Fabian
  Isensee}, \bibinfo{person}{Tristan~Anselm Kuder},
  \bibinfo{person}{Heinz-Peter Schlemmer}, {and} \bibinfo{person}{Klaus~H
  Maier-Hein}.} \bibinfo{year}{2020}\natexlab{}.
\newblock \showarticletitle{Retina U-Net: Embarrassingly simple exploitation of
  segmentation supervision for medical object detection}. In
  \bibinfo{booktitle}{\emph{Machine Learning for Health Workshop}}. PMLR,
  \bibinfo{pages}{171--183}.
\newblock


\bibitem[\protect\citeauthoryear{Liu, Cabahug-Zuckerman, Stubbs, Pendola, Cai,
  Mann, and Castillo}{Liu et~al\mbox{.}}{2019}]%
        {liu2019mechanical}
\bibfield{author}{\bibinfo{person}{Chao Liu}, \bibinfo{person}{Pamela
  Cabahug-Zuckerman}, \bibinfo{person}{Christopher Stubbs},
  \bibinfo{person}{Martin Pendola}, \bibinfo{person}{Cinyee Cai},
  \bibinfo{person}{Kenneth~A Mann}, {and} \bibinfo{person}{Alesha~B Castillo}.}
  \bibinfo{year}{2019}\natexlab{}.
\newblock \showarticletitle{Mechanical loading promotes the expansion of
  primitive osteoprogenitors and organizes matrix and vascular morphology in
  long bone defects}.
\newblock \bibinfo{journal}{\emph{Journal of Bone and Mineral Research}}
  \bibinfo{volume}{34}, \bibinfo{number}{5} (\bibinfo{year}{2019}),
  \bibinfo{pages}{896--910}.
\newblock


\bibitem[\protect\citeauthoryear{Long, Shelhamer, and Darrell}{Long
  et~al\mbox{.}}{2015}]%
        {long2015fully}
\bibfield{author}{\bibinfo{person}{Jonathan Long}, \bibinfo{person}{Evan
  Shelhamer}, {and} \bibinfo{person}{Trevor Darrell}.}
  \bibinfo{year}{2015}\natexlab{}.
\newblock \showarticletitle{Fully convolutional networks for semantic
  segmentation}. In \bibinfo{booktitle}{\emph{Proceedings of the IEEE
  conference on computer vision and pattern recognition}}.
  \bibinfo{pages}{3431--3440}.
\newblock


\bibitem[\protect\citeauthoryear{Mitchell and Schoen}{Mitchell and
  Schoen}{2010}]%
        {bloodVessels}
\bibfield{author}{\bibinfo{person}{Richard~N Mitchell} {and}
  \bibinfo{person}{Frederick~J Schoen}.} \bibinfo{year}{2010}\natexlab{}.
\newblock \showarticletitle{Blood vessels}.
\newblock \bibinfo{journal}{\emph{Robbins and Cotran: Pathologic Basis of
  Disease.(8th edition) Saunders Elsevier, Philadelphia, US}}
  (\bibinfo{year}{2010}), \bibinfo{pages}{516--17}.
\newblock


\bibitem[\protect\citeauthoryear{Moccia, De~Momi, El~Hadji, and Mattos}{Moccia
  et~al\mbox{.}}{2018}]%
        {moccia2018blood}
\bibfield{author}{\bibinfo{person}{Sara Moccia}, \bibinfo{person}{Elena
  De~Momi}, \bibinfo{person}{Sara El~Hadji}, {and} \bibinfo{person}{Leonardo~S
  Mattos}.} \bibinfo{year}{2018}\natexlab{}.
\newblock \showarticletitle{Blood vessel segmentation algorithms—review of
  methods, datasets and evaluation metrics}.
\newblock \bibinfo{journal}{\emph{Computer methods and programs in
  biomedicine}}  \bibinfo{volume}{158} (\bibinfo{year}{2018}),
  \bibinfo{pages}{71--91}.
\newblock


\bibitem[\protect\citeauthoryear{Ramasamy, Kusumbe, Wang, and Adams}{Ramasamy
  et~al\mbox{.}}{2014}]%
        {ramasamy2014endothelial}
\bibfield{author}{\bibinfo{person}{Saravana~K Ramasamy},
  \bibinfo{person}{Anjali~P Kusumbe}, \bibinfo{person}{Lin Wang}, {and}
  \bibinfo{person}{Ralf~H Adams}.} \bibinfo{year}{2014}\natexlab{}.
\newblock \showarticletitle{Endothelial Notch activity promotes angiogenesis
  and osteogenesis in bone}.
\newblock \bibinfo{journal}{\emph{Nature}} \bibinfo{volume}{507},
  \bibinfo{number}{7492} (\bibinfo{year}{2014}), \bibinfo{pages}{376--380}.
\newblock


\bibitem[\protect\citeauthoryear{Ronneberger, Fischer, and Brox}{Ronneberger
  et~al\mbox{.}}{2015}]%
        {U-Net}
\bibfield{author}{\bibinfo{person}{Olaf Ronneberger}, \bibinfo{person}{Philipp
  Fischer}, {and} \bibinfo{person}{Thomas Brox}.}
  \bibinfo{year}{2015}\natexlab{}.
\newblock \showarticletitle{U-Net: Convolutional Networks for Biomedical Image
  Segmentation}.
\newblock \bibinfo{journal}{\emph{CoRR}}  \bibinfo{volume}{abs/1505.04597}
  (\bibinfo{year}{2015}).
\newblock
\showeprint[arxiv]{1505.04597}
\urldef\tempurl%
\url{http://arxiv.org/abs/1505.04597}
\showURL{%
\tempurl}


\bibitem[\protect\citeauthoryear{Roth, Oda, Zhou, Shimizu, Yang, Hayashi, Oda,
  Fujiwara, Misawa, and Mori}{Roth et~al\mbox{.}}{2018}]%
        {ROTH201890}
\bibfield{author}{\bibinfo{person}{Holger~R. Roth}, \bibinfo{person}{Hirohisa
  Oda}, \bibinfo{person}{Xiangrong Zhou}, \bibinfo{person}{Natsuki Shimizu},
  \bibinfo{person}{Ying Yang}, \bibinfo{person}{Yuichiro Hayashi},
  \bibinfo{person}{Masahiro Oda}, \bibinfo{person}{Michitaka Fujiwara},
  \bibinfo{person}{Kazunari Misawa}, {and} \bibinfo{person}{Kensaku Mori}.}
  \bibinfo{year}{2018}\natexlab{}.
\newblock \showarticletitle{An application of cascaded 3D fully convolutional
  networks for medical image segmentation}.
\newblock \bibinfo{journal}{\emph{Computerized Medical Imaging and Graphics}}
  \bibinfo{volume}{66} (\bibinfo{year}{2018}), \bibinfo{pages}{90--99}.
\newblock
\showISSN{0895-6111}
\urldef\tempurl%
\url{https://doi.org/10.1016/j.compmedimag.2018.03.001}
\showDOI{\tempurl}


\bibitem[\protect\citeauthoryear{Saran, Piperni, and Chatterjee}{Saran
  et~al\mbox{.}}{2014}]%
        {saran2014role}
\bibfield{author}{\bibinfo{person}{Uttara Saran}, \bibinfo{person}{Sara~Gemini
  Piperni}, {and} \bibinfo{person}{Suvro Chatterjee}.}
  \bibinfo{year}{2014}\natexlab{}.
\newblock \showarticletitle{Role of angiogenesis in bone repair}.
\newblock \bibinfo{journal}{\emph{Archives of biochemistry and biophysics}}
  \bibinfo{volume}{561} (\bibinfo{year}{2014}), \bibinfo{pages}{109--117}.
\newblock


\bibitem[\protect\citeauthoryear{Sivaraj and Adams}{Sivaraj and Adams}{2016}]%
        {sivaraj2016blood}
\bibfield{author}{\bibinfo{person}{Kishor~K Sivaraj} {and}
  \bibinfo{person}{Ralf~H Adams}.} \bibinfo{year}{2016}\natexlab{}.
\newblock \showarticletitle{Blood vessel formation and function in bone}.
\newblock \bibinfo{journal}{\emph{Development}} \bibinfo{volume}{143},
  \bibinfo{number}{15} (\bibinfo{year}{2016}), \bibinfo{pages}{2706--2715}.
\newblock


\bibitem[\protect\citeauthoryear{Staal, Abr{\`a}moff, Niemeijer, Viergever, and
  Van~Ginneken}{Staal et~al\mbox{.}}{2004}]%
        {1282003}
\bibfield{author}{\bibinfo{person}{Joes Staal}, \bibinfo{person}{Michael~D
  Abr{\`a}moff}, \bibinfo{person}{Meindert Niemeijer}, \bibinfo{person}{Max~A
  Viergever}, {and} \bibinfo{person}{Bram Van~Ginneken}.}
  \bibinfo{year}{2004}\natexlab{}.
\newblock \showarticletitle{Ridge-based vessel segmentation in color images of
  the retina}.
\newblock \bibinfo{journal}{\emph{IEEE Transactions on Medical Imaging}}
  \bibinfo{volume}{23}, \bibinfo{number}{4} (\bibinfo{year}{2004}),
  \bibinfo{pages}{501--509}.
\newblock
\urldef\tempurl%
\url{https://doi.org/10.1109/TMI.2004.825627}
\showDOI{\tempurl}


\bibitem[\protect\citeauthoryear{Sun, Wang, Fu, Li, Liu, and Zhao}{Sun
  et~al\mbox{.}}{2022}]%
        {sun2022machine}
\bibfield{author}{\bibinfo{person}{Mingzhu Sun}, \bibinfo{person}{Yiwen Wang},
  \bibinfo{person}{Zhenhua Fu}, \bibinfo{person}{Lu Li},
  \bibinfo{person}{Yaowei Liu}, {and} \bibinfo{person}{Xin Zhao}.}
  \bibinfo{year}{2022}\natexlab{}.
\newblock \showarticletitle{A Machine Learning Method for Automated In Vivo
  Transparent Vessel Segmentation and Identification Based on Blood Flow
  Characteristics}.
\newblock \bibinfo{journal}{\emph{Microscopy and Microanalysis}}
  (\bibinfo{year}{2022}), \bibinfo{pages}{1--14}.
\newblock


\bibitem[\protect\citeauthoryear{Tang, Zou, Yang, Shi, Dan, and Song}{Tang
  et~al\mbox{.}}{2020}]%
        {tang2020two}
\bibfield{author}{\bibinfo{person}{Wei Tang}, \bibinfo{person}{Dongsheng Zou},
  \bibinfo{person}{Su Yang}, \bibinfo{person}{Jing Shi},
  \bibinfo{person}{Jingpei Dan}, {and} \bibinfo{person}{Guowu Song}.}
  \bibinfo{year}{2020}\natexlab{}.
\newblock \showarticletitle{A two-stage approach for automatic liver
  segmentation with Faster R-CNN and DeepLab}.
\newblock \bibinfo{journal}{\emph{Neural Computing and Applications}}
  \bibinfo{volume}{32}, \bibinfo{number}{11} (\bibinfo{year}{2020}),
  \bibinfo{pages}{6769--6778}.
\newblock


\bibitem[\protect\citeauthoryear{Voulodimos, Doulamis, Doulamis, and
  Protopapadakis}{Voulodimos et~al\mbox{.}}{2018}]%
        {voulodimos2018deep}
\bibfield{author}{\bibinfo{person}{Athanasios Voulodimos},
  \bibinfo{person}{Nikolaos Doulamis}, \bibinfo{person}{Anastasios Doulamis},
  {and} \bibinfo{person}{Eftychios Protopapadakis}.}
  \bibinfo{year}{2018}\natexlab{}.
\newblock \showarticletitle{Deep learning for computer vision: A brief review}.
\newblock \bibinfo{journal}{\emph{Computational intelligence and neuroscience}}
   \bibinfo{volume}{2018} (\bibinfo{year}{2018}).
\newblock


\bibitem[\protect\citeauthoryear{Zhou, Rahman~Siddiquee, Tajbakhsh, and
  Liang}{Zhou et~al\mbox{.}}{2018}]%
        {zhou2018unet++}
\bibfield{author}{\bibinfo{person}{Zongwei Zhou}, \bibinfo{person}{Md~Mahfuzur
  Rahman~Siddiquee}, \bibinfo{person}{Nima Tajbakhsh}, {and}
  \bibinfo{person}{Jianming Liang}.} \bibinfo{year}{2018}\natexlab{}.
\newblock \showarticletitle{Unet++: A nested u-net architecture for medical
  image segmentation}.
\newblock In \bibinfo{booktitle}{\emph{Deep learning in medical image analysis
  and multimodal learning for clinical decision support}}.
  \bibinfo{publisher}{Springer}, \bibinfo{pages}{3--11}.
\newblock


\bibitem[\protect\citeauthoryear{Zhou, Siddiquee, Tajbakhsh, and Liang}{Zhou
  et~al\mbox{.}}{2019}]%
        {U-Net++v2}
\bibfield{author}{\bibinfo{person}{Zongwei Zhou},
  \bibinfo{person}{Md~Mahfuzur~Rahman Siddiquee}, \bibinfo{person}{Nima
  Tajbakhsh}, {and} \bibinfo{person}{Jianming Liang}.}
  \bibinfo{year}{2019}\natexlab{}.
\newblock \showarticletitle{Unet++: Redesigning skip connections to exploit
  multiscale features in image segmentation}.
\newblock \bibinfo{journal}{\emph{IEEE transactions on medical imaging}}
  \bibinfo{volume}{39}, \bibinfo{number}{6} (\bibinfo{year}{2019}),
  \bibinfo{pages}{1856--1867}.
\newblock


\end{thebibliography}










\end{document}